\documentclass[prl,aps,twocolumn,showpacs,amsmath,amssymb]{revtex4}

\usepackage{graphicx}
\usepackage{dcolumn}
\usepackage{bm}

\begin{document}
\title{In-plane ferromagnetism in charge-ordering $Na_{0.55}CoO_2$ }
\author{ C. H. Wang$^1$}
\author{X. H. Chen$^1$}
\altaffiliation{Corresponding author} \email{chenxh@ustc.edu.cn}
\author{T. Wu$^1$, X. G. Luo$^1$, G. Y. Wang$^1$, J. L. Luo$^2$ }
\affiliation{1. Hefei National Laboratory for Physical Science at
Microscale and Department of Physics, University of Science and
Technology of China, Hefei, Anhui 230026, People's Republic of
China\\ 2. Beijing National Laboratory for Condensed Matter
Physics, Institute of Physics, Chinese Academy of Sciences,
Beijing 100080, China}

\date{\today}

\begin{abstract}

The magnetic and transport properties are systematically studied
on the single crystal $Na_{0.55}CoO_2$ with charge ordering and
divergency in resistivity below 50 K. A long-range ferromagnetic
ordering is observed in susceptibility below 20 K with the
magnetic field parallel to Co-O plane, while a negligible behavior
is observed with the field perpendicular to the Co-O plane. It
definitely gives a direct evidence for the existence of in-plane
ferromagnetism below 20 K. The observed magnetoresistance (MR) of
30 $\%$ at the field of 6 T at low temperatures indicates an
unexpectedly strong spin-charge coupling in triangle lattice
systems.
\end{abstract}


\maketitle
\newpage

Discovery of superconductivity with $T_c\sim$ 5 K in
$Na_{0.35}CoO_2\cdot1.3H_2O$ \cite{Takada1} makes one consider
that $Na_xCoO_2$ may be a good example like cuprates as a doped
Mott insulator becomes superconducting. A characteristic of
non-hydrate $Na_xCoO_2$ is the sensitivity of the electronic
states to slight change in x. A charge ordering state with
anomalous change in thermopower, Hall coefficient and thermal
conductivity occurs in the sample $Na_{0.5}CoO_2$ with Co valence
of +3.5.\cite{foo}

Takada et al. have reinvestigated the superconducting sodium
cobalt oxide bilayer-hydrate and revised the stoichiometry of the
superconductor to be $Na_{0.337}(H_3O)_{0.234}CoO_2\cdot
yH_2O$.\cite{takada2} It is found that the oxonium ions,
$(H_3O)^+$, can occupy the same crystallographic sites as the Na
ions when the sample of the $Na_{0.4}CoO_2$ is immersed in
distilled water.\cite{takada2} The occupation of oxonium ions in
the $Na^+$ layers in superconducting hydrate $Na_xCoO_2$ makes the
Co valence to be +3.4, being much lower than +3.7 which is
directly deduced from the Na content and is widely used to discuss
the superconductivity, and to draw superconducting phase diagram
and to make theoretical
calculations.\cite{foo,singh,Kumar,honerkamp,Baskaran1} Recently
Barnes et al. mapped a dome-shaped superconducting phase diagram
as a function of cobalt oxidation state. They found that the
maximum $T_c$ in $Na_xCoO_2 \cdot yH_2O$
 occurs when cobalt oxidation state is near +3.5.\cite{barnes}
 The maximum  $T_c$ occurs near the charge ordered insulating
 state that correlates with the average cobalt oxidation of +3.5.
  Sakurai et al. also gave a superconducting  phase diagram with a
  cobalt valence of +3.48.\cite{sakurai} They found that the superconductivity
  was significantly affected by the isovalent exchange of $Na^+$ and $H_3O^+$,
  rather than by variation
  of Co valence. In addition, the superconducting phase has been revised by
 different groups.\cite{milne,sakurai,sakurai1,ihara} It is found
 that the superconductivity with the highest $T_c$ is observed in
 the vicinity of a magnetic phase, suggesting strongly that
 magnetic fluctuations play an important role in the occurrence of
superconductivity.\cite{sakurai1,ihara} Based on the above
results, the superconductivity occurs in the cobalt oxidation
state near +3.5 instead of about +3.7. Therefore, one should take
the $Na_xCoO_2$ with $x \sim 0.5$ as the parent compound for the
superconductor. It suggests that the superconducting state could
compete with the charge ordering state. This assumption could be
supported by the abrupt upturn at $T^*\sim 52$ K before
superconducting transition in both the ab plane and the c
direction resistivity.\cite{jin} Therefore, it is very significant
to study the magnetic properties of $Na_xCo_2$ with x around 0.5
for understanding the pairing mechanism.

We systematically study transport and magnetic properties for the
$Na_xCoO_2$ crystals with x  around 0.5. It  is found that the
single crystals show paramagnetic metallic behavior with x
slightly less than 0.5 (cobalt oxidation of 3.5), while an
insulating behavior below 50  K in resistivity and a complicated
magnetic properties with x slightly larger than 0.5. In this
 letter, we present the transport and magnetic properties in
 $Na_{0.55}CO_2$ with cobalt oxidation of 3.45. An in-plane
 ferromagnetism is observed below 20 K which depends on x. A large MR
of 30 \% with field of  6 T at low temperatures are surprisingly
sensitive to spin orientation, suggesting that charge transport is
strongly coupled to spin. If superconductivity really occurs in
the  cobalt oxidation near 3.5, the strong spin-charge coupling
should be considered for the superconducting mechanism.

High quality single crystals $Na_{0.7}CoO_2$ were grown using the
flux method. The typical dimensional is about $2 \times 1.5 \times
0.01 mm^3$ with the shortest dimension along the c axis.  The
$Na_{0.55}CoO_2$ sample is prepared by sodium de-intercalation
from the $Na_{0.7}CoO_2$ singe crystals. The procedure is similar
to the way to prepare the $Na_{0.5}CoO_2$ sample, the difference
is that the lower $I_2$ concentration is used with the shorter
reaction time compared to the preparation of $Na_{0.5}CoO_2$.
About 5 mg $Na_{0.7}CoO_2$ single crystals were immersed in the
sealed conical flask with 5ml 0.15 M solution of $I_2$ in
acetonitrile at room temperature for about 30 hours. The actual Na
concentration was determined by Atomscan Advantage inductively
coupled plasma atomic emission spectrometer (ICP). Both the
resistance and magnetic susceptibility were measured with the
magnetic field parallel and perpendicular to the Co-O plane. The
resistivity and magnetoresistance were performed in Quantum Design
PPMS systems. The magnetic properties were measured with a Quantum
 Design SQUID magnetometer. It should be addressed that the samples
 and all results discussed as follow are well reproducible.

Figure 1(a) and 1(b) show the zero-field cooled (ZFC) and
field-cooled (FC) magnetic susceptibility $\chi$ as a function of
temperature for single crystal $Na_{0.55}CoO_2$ in different
fields (H) with H $\|$ Co-O plane and H $\bot$ Co-O plane,
respectively. As shown in Fig.1, the sample shows a Pauli
paramagnetic behavior in the high temperature range, being similar
to that of the sample $Na_{0.5}CoO_2$.\cite{foo} However, the
sample $Na_{0.55}CoO_2$ exhibits only one anomaly at $T_c \sim 77$
K in $\chi$ taken for the applied magnetic field parallel to the
ab-plane, and no anomaly is observed in $\chi$ taken the applied
magnetic field perpendicular to the ab-plane. This result is
consistent with the antiferromagnetic ordering at $T_{c1}\sim 87$
K with spin direction within the ab-plane in
\begin{figure}[h]
\includegraphics[width=9cm]{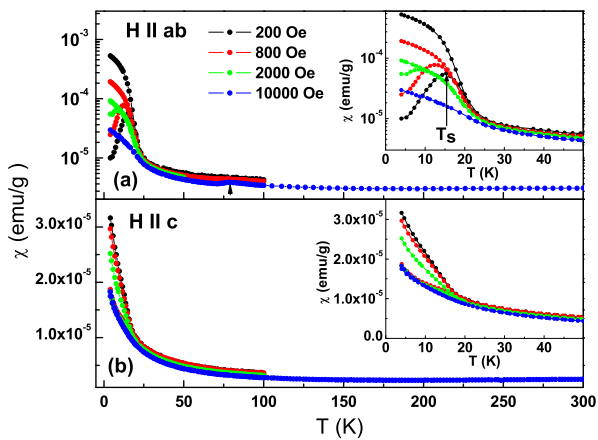}\vspace{-5mm}
\caption{\label{fig:epsart} ZFC and FC susceptibility $\chi$ as a
function of temperature for single crystal $Na_{0.55}CoO_2$ in
different fields of 200, 800, 2000 and 10000 Oe (H) with (a): H
$\|$ Co-O plane and  (b) H $\bot$ Co-O plane, respectively. In
order to clearly show the variation of susceptibility with field
in the low temperatures, the data below 30 K are shown in the
inset. }
\end{figure}
$Na_{0.5}CoO_2$.\cite{yokoi} No the anomaly at $T_{c1}\sim 53$ K,
which appears in $\chi$ of $Na_{0.5}CoO_2$, is observed in $\chi$
of $Na_{0.55}CoO_2$, and the $T_{c1}$ decreases by 10 K compared
to $Na_{0.5}CoO_2$. It suggests that the $T_{c1}$ and $T_{c2}$ is
very sensitive to the Na content. A striking feature is that a
ferromagnetic ordering occurs blow 20 K. The ZFC and FC branches
show up in the low H, while the ZFC and FC data are the same with
the H larger than 1 T. It suggests a weak ferromagnetic behavior.
For the case of the applied magnetic field parallel to ab-plane,
there exists a maximum in ZFC $\chi$ at a certain temperature
$T_s$ below the ferromagnetic transition temperature ($T_c$). As
shown in the inset of Fig.1(a), the $T_s$ decreases with
increasing the applied field. This is because the applied field
makes the spin frozen temperature to decrease. Compared to the
case of H parallel to ab-plane, the $\chi$ slightly increases with
decreasing the temperature and the FC $\chi$ for different fields
is almost the same  with the applied field perpendicular to
ab-plane below the $T_c$. In addition, the $\chi$ for H$\|$ Co-O
plane is much larger (more than one order) than that for  H $\bot$
Co-O plane. It suggests that the spins align ferromagnetically
with their direction with the ab-plane.

 To confirm the in-plane ferromagnetic ordering, the magnetic hysteresis (M-H)
 is studied with the applied field parallel to and perpendicular
 to ab-plane. Figure 2(a) shows a M-H loop at 4 K with the
magnetic field  H $\|$ Co-O plane. It suggests that there exists a
ferromagnetic ordering in the low temperature. But the magnetic
moment does not saturate and increases linearly with H in the high
fields. The linear component could arise from the in-plane
\begin{figure}[h]
\includegraphics[width=9cm]{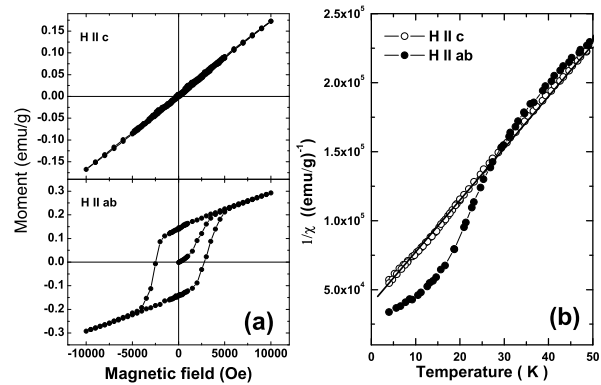}\vspace{-5mm}
\caption{\label{fig:epsart} (a): M-H loops at 4 K with the
magnetic field  H $\|$ Co-O plane and  H $\bot$ Co-O plane,
respectively. (b): To clearly show the anisotropic
susceptibility, the inverse susceptibility as a function of
temperature below 50 K at H=10000 Oe parallel to and perpendicular
to ab plane, respectively. }
\end{figure}
antiferromagnetic ordering at $T_{c1}\sim 77$ K. For the case of H
$\bot$ Co-O plane, no M-loop is observed and magnetic moment
increases linearly with increasing H, indicating a paramagnetic
behavior. These results clearly indicate that the ferromagnetic
ordering observed in Fig.1 occurs with spin direction within the
ab-plane. In order to compare the in-plane susceptibility with
out-of-plane susceptibility, the inverse susceptibility $\chi$
with applied magnetic field H $\|$ Co-O plane and H $\bot$ Co-O
plane as a function of temperature is shown in Fig.2(b) at H=10000
Oe below 50 K. It indicates that the susceptibility with H $\bot$
Co-O plane shows a Curie-Weiss behavior below 50 K, while a weak
ferromagnetic ordering occurs in $\chi$ with H $\|$ Co-O plane.
This result is consistent with that observed in Fig.2(a).

To further study the effect of the anisotropic susceptibility on
the charge transport and the coupling between charge and spin, the
charge transport is systematically studied. Figure 3 shows the
in-plane resistivity and magnetoresistance with H $\bot$ Co-O
plane and H $\|$ Co-O plane. As shown in the inset of Fig.3(a),
the zero-field resistivity shows nearly the same behavior as that
of the charge ordering $Na_{0.5}CoO_2$.\cite{foo} However, a
transition of the slope $\emph{d}\rho/\emph{d}T$ from negative to
positive is observed at about 17 K ($T_{\rho}$), which coincides
with the ferromagnetic transition temperature shown in Fig.1. It
suggests that such transition arises from the ferromagnetic
ordering.
\begin{figure}[h]
\includegraphics[width=9cm]{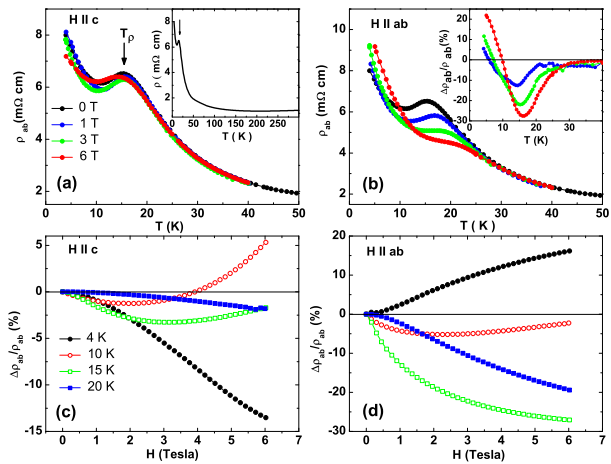}\vspace{-5mm}
\caption{\label{fig:epsart} Temperature dependence of in-plane
resistivity under different fields with (a):H $\bot$ Co-O plane
and (b): H $\|$ Co-O plane; Isothermal magnetoresistance with (c):
H $\bot$ Co-O plane and (d): H $\|$ Co-O plane. }
\end{figure}

The in-plane resistivity $\rho_{ab}$ as a function of temperature
at H=1, 3 and 6 T with  H $\|$ Co-O plane and H $\bot$ Co-O plane
is shown in Fig.3(a) and (b), respectively. It is found that the
magnetic field has a remarkable effect on $\rho_{ab}$  below 25 K,
while a negligible effect on $\rho_{ab}$ above 25 K. The effect of
magnetic field on $\rho_{ab}$ with H $\|$ Co-O plane is much
larger than that with H $\bot$ Co-O plane around the transition
temperature ($T_{\rho}$). As shown in Fig.3(b), a large negative
magnetoresistance (MR) is observed around $T_{\rho}$ with H $\|$
Co-O plane, and the $T_{\rho}$ increases with increasing H. For
the case of H $\bot$ Co-O plane, the MR is much less than that
with H $\|$ ab plane around $T_{\rho}$ and the $T_{\rho}$ is
nearly independent on the field. These results can be well
understood by the in-plane ferromagnetic ordering. This is because
the in-plane ferromagnetic ordering is not affected by the H
perpendicular to ab-plane.  As shown in the inset of Fig.3(b), a
maximum negative MR shows up abound $T_{\rho}$ with H $\|$ Co-O
plane, while a positive MR occurs at low temperatures. The
isothermal MR with H $\bot$ Co-O plane and H $\|$ Co-O plane
 at 4, 10, 15, and 20 K is shown in Fig.3(c) and (d), respectively. For the case
of  H $\bot$ Co-O plane, the isothermal MR shows a complicated
behavior. At 4 and 20 K, the MR is negative and increases
monotonically  with increasing H, while at 15 K the MR is negative
and shows a maximum at $H\sim 3$ T; at 10 K a negative maximum in
the MR is observed at  $H\sim 2$ T, but with further increasing H
the MR sign changes from negative to positive. It suggests that
the isothermal MR at 10 and 15 K consists of two contributions:
one \emph{negative} and one \emph{positive} component. For the
case of H $\|$ Co-O plane,  at 15 K and 20 K the isothermal MR is
negative and increases monotonously with H, and the MR is as high
as $\sim$30\% at H=6 T and 15 K just below $T_{\rho}$. The large
negative MR at 20 K arises from the enhancement of the
ferromagnetic transition temperature induced by H. In contrast to
the H $\bot$ Co-O plane, at 4 K the MR is positive and increases
with increasing H. At 10 K, the MR shows a similar behavior to
that with H $\bot$ Co-O plane. These results definitely indicate
that there exist two contributions (one \emph{negative} and one
\emph{positive} component) to the MR. The large negative component
should arise from the ferromagnetic ordering. In contrast to the
case of H $\|$ Co-O plane, a negative MR is observed at low
temperatures for H $\bot$ Co-O plane. Similar phenomena have been
observed in $Na_{0.5}CoO_2$.\cite{wang} It suggests that the low
temperature MR could be dominated by the antiferromagnetic
ordering.

The out-of-plane resistivity $\rho_c$ under H and the isothermal
MR are also studied. Figure 4 shows the c-axis magnetotransport
and the isothermal MR with  H $\|$ Co-O plane and H $\bot$ Co-O
plane, respectively.
\begin{figure}[h]
\includegraphics[width=8cm]{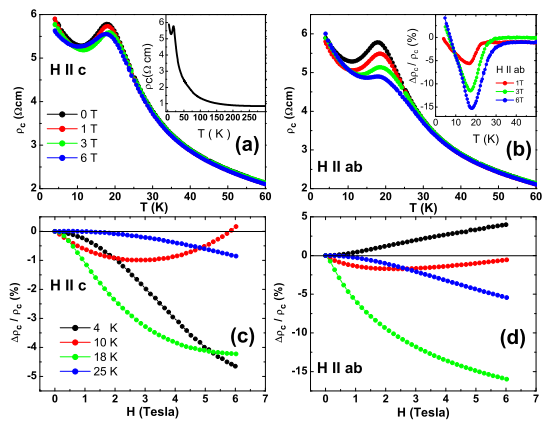}
\caption{\label{fig:epsart} Temperature dependence of out-of-plane
resistivity under different fields with (a):H $\bot$ Co-O plane
and (b): H $\|$ Co-O plane; Isothermal magnetoresistance with (c):
H $\bot$ Co-O plane and (d): H $\|$ Co-O plane.}
\end{figure}
The effect of magnetic field on $\rho_c$ is quite similar to that
on $\rho_{ab}$ although the out-of-plane MR is less than the
in-plane MR. Similar to in-plane behavior, the $T_{\rho}$ is not
affected by H applied along c-axis, and is enhanced by H parallel
to the ab plane. These results further support that ferromagnetic
ordering occurs with spin direction within Co-O plane. In
addition, the MR with H$\|$ ab plane is negative at low
temperatures, while the MR with $\bot$ ab plane.

NMR and neutron diffraction studies have given spin structure for
the charge ordering $Na_{0.5}CoO_2$.\cite{yokoi,gasparovic} They
reported that there exist two kinds of Co sites with large and
small magnetic moments in $Na_{0.5}CoO_2$. As shown in Fig.5, the
large magnetic moments align antiferromagnetically at $T_{c1}\sim
87$ K with their direction within the ab plane, while the small
magnetic moments align along the direction parallel to the c-axis.
It cannot be distinguish if the in-plane spin correlation of the
small moment sites is ferromagnetic or anitferromagnetic. The
splitting of the zero-field NMR signals from the Co sites with
small moment has been observed at $T_{c2}\sim 53$ K. It suggests
that the transition at $T_{c2}\sim 53$ K arises from the spin
correlation of the small magnetic moment.
\begin{figure}[h]
\includegraphics[width=8cm]{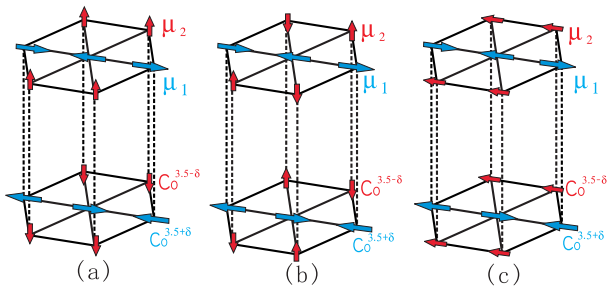}
\caption{\label{fig:epsart} (a) and (b):The magnetic structures
proposed for $Na_{0.5}CoO_2$ from Ref.14; (c) The magnetic
structure for $Na_{0.55}CoO_2$.}
\end{figure}
 As shown in Fig.1, the anomaly at $T_{c2}\sim 53$ K is not
 observed, while the antiferromagnetic transition at $T_{c1}$
 still occurs in $Na_{0.55}CoO_2$. It implies that the in-plane ferromagnetic
  ordering  with the spin direction within Co-O plane should originate from
  the Co sites with small magnetic ordering.
Therefore, the Co sites with large magnetic moment align
 antiferromagnetically at $T_{c1}\sim 77$ K with the spin
 direction within the plane, similar to that in
 $Na_{0.5}CoO_2$, while the Co sites with small magnetic moment
 couple ferromagnetically at $T_{c}\sim 17$ K with the spin
 alignment in Co-O plane. A spin structure shown in Fig.5(c) is proposed for
 $Na_{0.55}CoO_2$. The neutron scattering studies on
 $Na_{0.82}CoO_2$\cite{Bayrakci} and $Na_{0.75}CoO_2$\cite{Boothroyd,Helme} indicate that the in-plane
 and inter-plane spin correlation is ferromagnetic and
 antiferromagnetic, respectively, and the spins align along the
 c-axis. A spin-flop takes place with decreasing Na, that is: the
 spin direction changes from along c-axis to within Co-O plane.
 For the $Na_{0.5}CoO_2$, the small magnetic moment of the Co
 sites still is  along c-axis, while the large magnetic moments of the
Co sites align in ab-plane. Compared to the $Na_{0.5}CoO_2$, the
spin-flop occurs and the spin direction of the Co sites with small
magnetic moment changes from along c-axis to in Co-O plane in
$Na_{0.55}CoO_2$. It indicates that the spin structure is
sensitive to Na content. From the point of effective carrier
concentration, the Co oxidation state is +3.45 in
$Na_{0.55}CoO_2$, being close to +3.48 in the superconductor
$Na_{0.337}(H_3O)_{0.234}CoO_2\cdot yH_2O$. Therefore, we give a
direct evidence that there exists a ferromagnetic state in the
vicinity of the superconducting state. It further confirms  the
results reported by Sakurai and co-workers.\cite{sakurai,ihara}
The ferromagnetic ordering with spin direction within Co-plane
indicates that it is possible for spin-triplet superconductivity.
Indeed, a theoretical calculation has predicted spin-triplet
superconductivity with $T_c$ negligibly reduced by the magnetic
field.\cite{yanase}

In conclusion, the magnetic properties and magnetotransport are
systematically studied in $Na_{0.55}CoO_2$. The results of
susceptibility and magnetotransport definitely indicate an
in-plane ferromagnetic ordering below 20 K. In Co-O plane, there
exist an antiferromagnetic ordering from the Co-sites with large
magnetic moment and a ferromagnetic ordering from the Co-sites
with small magnetic moment, and their spin direction is within
Co-O plane. The Co oxidation state of +3.45 in $Na_{0.55}CoO_2$ is
close to +3.48 in the superconductor
$Na_{0.337}(H_3O)_{0.234}CoO_2\cdot yH_2O$, suggesting that it is
possible for spin-triplet superconductivity in Co-triangle lattice
system.

 \vspace*{-2mm}
\section{Acknowledgment}
This work is supported by the Nature Science Foundation of China
and by the Ministry of Science and Technology of China, and by the
Knowledge Innovation Project of Chinese Academy of Sciences.

\end{document}